\documentclass[rnote]{aa} 
\usepackage{graphicx}
\usepackage{txfonts}
\begin{document}
   \title{ An optical search for supernova remnants in the nearby spiral galaxy NGC 2903}

   \author{E. Sonbas
          \inst{1}
         \fnmsep\inst{2} 
           \and
           A. Akyuz \inst{1}
          \and
          S. Balman \inst{3}
        }
      
   \offprints{E. Sonbas}

   \institute{University of Cukurova, Department of Physics, 01330 Adana, Turkey
             \and
	     Special Astrophysical Observatory of R.A.S., Karachai-Cherkessia,
              Nihnij Arkhyz, 369167 Russia
	     \and
              Dept. of Physics, Middle East Technical University, 06531 Ankara, Turkey 
            }
	     
   \date{Received -; accepted -}

 
  \abstract
   {}
   {We present the results of an optical search for supernova remnants (SNRs) in the 
   nearby spiral galaxy NGC 2903.}
   { Interference filter images and spectral data were taken in March 2005 with the f/7.7 1.5 m Russian Turkish Telescope (RTT150) at TUBITAK 
   National Observatory (TUG). Spectral
   data were obtained with the 6 m BTA (Bolshoi Azimuthal Telescope, Russia). 
   We used 
   the SNR identification criterion that consists of constructing the continuum-subtracted H$\alpha$ and 
   continuum-subtracted [SII]$\lambda$$\lambda$6716,6731 images and their ratios.}
   {Five SNR candidates were identified in NGC 2903 with [SII]/H$\alpha$ ratios ranging from 0.41 - 0.74 and 
    H$\alpha$ intensities ranging from 9.4$\times$10$^{-15}$ to 1.7$\times$10$^{-14}$ ergs cm$^{-2}$ s$^{-1}$.   
   This work represents the first identification of SNRs by an optical survey in NGC 2903.
 We present the spectrum of one of the bright candidates and derive an [SII]/H$\alpha$ emission line ratio of
 0.42 for this source. 
 In addition, the weak [OIII]$\lambda$5007/H$\beta$ emission line ratio in the spectrum of this SNR indicates  an old oxygen-deficient remnant with a low propagation velocity.}
   {}


   \maketitle
%

\section{Introduction}

   Supernova remnants (SNRs) are important for many of the theories of interstellar medium (ISM)
   because supernova explosions and their eventual dispersion of ejected material have the effect of 
   enriching the ISM with the material processed in stellar interiors. 
In a typical HII region, the sulfur is in the form of $S^{++}$
   because of the strong photoionization flux of the central hot star or stars. Therefore, 
   the [SII]/H$\alpha$ ratio is typically $\approx$ 0.1 - 0.3 for HII regions. 
   Outside the HII region, there are not
   enough energetic free electrons to excite $S^{+}$ and to produce forbidden-line ([SII]$\lambda\lambda6716,6731$) emission. 
   For that reason, nearly all discrete emission nebulae with [SII]/H$\alpha$
   $\geq$ 0.4 are shock-heated and they are probably SNRs.   

In several prior works, the motivation for observing SNRs in nearby galaxies (Matonick \& Fesen 1997 (\textit{hereafter} MF97); Pannuti et al. 2000, 2002) and in our Galaxy (Mavromatakis et al. 2002; Fesen et al. 1997, 2008) has already been discussed. The sample of Galactic SNRs is quite large, and
   interstellar extinction and uncertain distances cause selection effects. 
    These problems are much less significant in extragalactic samples. Assuming the distance to a galaxy is known, all the SNRs are at the same distance from us, so some properties can be compared directly. Also, the foreground extinction is generally low, thus relative positions of SNR samples are determined accurately in an extragalactic survey. By knowing  the positions of the SNRs, their distributions are calculated relative to HII regions and spiral arms. Possible SNR progenitors have been investigated from these distributions (MF97; Blair \& Long 1997, 2004).

   Extragalactic searches for SNRs were first obtained for the Magellanic Clouds  
   by Mathewson \& Clarke (1973). They were the first to use the [SII]/H$\alpha$ emission line
   ratios for optical identification of SNRs. Blair et al. (1981), Smith et al. (1993), and Blair \& Long (1997,2004) also used same method successfully. 

 A number of nearby spiral galaxies have already been observed to identify SNRs using optical 
   observations (e.g D$^\prime$Odorico, Dopita \& Benvenuti 1980; Braun \& Walterbos 1993; Magnier et al. 1995; 
   MF97; Matonick et al. 1997; Gordon et al. 1998, 1999; 
   Blair \& Long 1997, 2004) and  X-ray observations (Pence et al. 2001; Ghavamian et al. 2005). Radio searches for extragalactic SNRs have been 
   conducted by Lacey et al. 1997; Lacey \& Duric (2001) and Hyman et al. (2001). SNR surveys
   have also been carried out at optical, radio, and X-ray wavelengths by Pannuti et al. (2000, 2002,
   2007).

  In this work,  we searched for SNRs in  the nearby spiral galaxy NGC 2903 using 
  the criterion [SII]/H$\alpha$ ratio is $\geq$ 0.4. 
  NGC 2903 is an SB(s)d type galaxy with a $61^{\circ}$ inclination angle, 
  and $17^{\circ}$ position angle  at a distance of 9.4 Mpc to NGC 2903 has been adopted for this paper (Bresolin et
al. 2005). 
 It was observed at radio (Williams \& Becklin 1985; Tsai et al. 2006),
infrared (Simons et al. 1988; Williams \& Becklin 1985; Alonso - Herrero, Ryder \& Knapen 2001),
 X - ray (Fabbiano, Trinchieri \& MacDonald 1984; Mizuno et al. 1998; Junkes \& Hensler 1996; 
 Tschoke, Hensler \& Junkes 2003), and optical wavelengths (Bresolin et al. 2005).
  Multiwavelength observations of NGC 2903 implies that it has a very complex structure with knots
in the nucleus. The knots, called ``hot-spots'', contain many early type stars
(Oka et al. 1974; Simons et al. 1988). Tsai et al. (2006) report subarcsecond-resolution VLA (Very Large Array) imaging of NGC 2903.
    They found seven discrete radio sources in the central 15$\arcsec$ $\times$ 15$\arcsec$ of galaxy at
the wavelengths of 6 and 2 cm to a limiting integrated flux density of 0.2 and 0.3 mJy, respectively. 
They identified one of their sources (source D) as a candidate radio SNR.
Their detected sources meet at least one of the following criteria: 5$\sigma$ detection at the peak intensity at one
wavelength and  4$\sigma$ emission detection at both 6 and 2 cm.

   The organization of the paper may be described as follows:.
    In Sect. 2 we discuss both the imaging and spectroscopic observations that were conducted 
   as part of this study, as well as the accompanying data reduction. The SNR identification technique, search results, 
   and discussion are described in Sect. 3.  
 
     \section{Observations and data reduction}
     
     \subsection{Imaging}
     
     NGC 2903 was observed in 2006 March with the 1.5 m Russian Turkish Telescope (RTT150) at TUBITAK National Observatory (TUG) in Turkey.
     Images were taken with TFOSC (TUBITAK Faint Object
     Spectrograph and Camera) 2048 $\times$ 2048 pixel CCD with a plate scale of 0\arcsec.39 pixel$^{-1}$, giving 
     13\arcmin.3 $\times$ 13\arcmin.3 FOV. We used narrowband interference filters
     centered on the lines of [SII] \& H$\alpha$ + 2 continuum filters to remove starlight from
     H$\alpha$ and [SII] images. Characteristics of the interference filters used in these
     observations are listed in Table 1. An observation log of the
     imaging data is shown in Table 2 for this galaxy. 
     The data were reduced using ESO--MIDAS
     (The European Southern Observatory Munich Image Data Analysis System) software environment. 
     Several H$\alpha$,  [SII], and associated continuum images of each galaxy field were
     combined to  obtain deeper field images
      in order to  increase the signal-to-noise ratio for the faintest objects. As shown in Table 2, 20 exposures (5400 seconds in total for two filters)  were combined 
     for two observation nights. Standard stars 
     from the list of  Oke (1974) and Stone (1977) were observed each night to determine the flux conversion
     factors. Bias frames and dome flats were also observed. 
     Each exposure was bias-subtracted, trimmed, and
     flat-fielded. The cosmic rays were removed from each [SII] and H$\alpha$ image.
     The SNR candidates overlaid on DSS (Digitized Sky Survey) images of the spiral
     galaxy can be found in Fig. 1. 
     
     \subsection{Spectroscopy} 
     
     The spectral data of one bright SNR were obtained with the optical telescope 
     BTA-6 m (Bolshoi Azimuthal Telescope, Russia) 2008 April. The SCORPIO (Spectral Camera with Optical Reducer
     for Photometrical and Interferometrical Observations) spectrograph was
     used in BTA with a CCD 2048 $\times$ 2048 pixels in size. We used 1$\arcsec$ slit width, and a 3500 - 7200 $\AA$ spectral range was assumed for SCORPIO with 10 $\AA$
     spectral resolution. The IDL codes and  IRAF packages were used to perform the basic reductions, flux, and
     wavelength calibrations and interstellar extinction correction. 
     Spectrophotometric standard stars from Oke (1974) and
     Stone (1977) catalogs were observed each night. Derived fluxes from
     the spectrophotometric standards were used to find the fluxes for spectral lines in our SNR spectrum. Biases, halogen  lamp
     flats, and FeAr or Neon calibration lamb exposures were obtained for each observation set. Finally  [SII] and
     H$\alpha$ emission line fluxes were measured using \textit{splot} routine in IRAF.   To obtain 
     spectral data of SNRs, the slit position was arranged so that both a bright star and the SNR were inside the slit. 
 The aim of spectral observations was to resolve [SII] $\lambda\lambda 6716, 6731$ doublet lines and to extract and eliminate contributions
     of two [NII] $\lambda\lambda 6548, 6583$ lines near the H$\alpha$ line.

     \section{Results and discussion}
     
     To identify the SNR candidates in NGC 2903, 
     we used the
     technique where continuum-subtracted H$\alpha$ and
     [SII] $\lambda\lambda 6716, 6731$ images were constructed and then 
  [SII]/ H$\alpha$ image ratios were made. Finally, regions that have image ratio values 
   $\geq$ 0.4 were identified as candidate SNRs (Blair \& Long 1997).
     
     Preliminary SNR candidates were found by blinking between continuum-subtracted [SII] and H$\alpha$ subfield images.
     We only displayed a region
     about 2$\arcmin$ on a side at one time for visual inspection of the fields to search for
     candidates. Any bright feature in the continuum-subtracted [SII] image was
     checked against the continuum-subtracted H$\alpha$ to make sure the 
     stars were not
     poorly subtracted. If the feature in the [SII] image 
     looked brighter than it was
     in the H$\alpha$ image, we marked it as an
     SNR candidate. Each preliminary SNR candidate was a possible target for
     follow-up spectral observation. In the latter step [SII] and H$\alpha$, 
     continuum-subtracted images were used to measure the total counts for each
     SNR candidate. We selected 
     a circular aperture in continuum-subtracted images to sum the ADU (Analogue to Digital Units) counts.
  Afterwards a concentric annulus was selected to determine the background counts to subtract from 
     the aperture sum. The aperture sizes used to measure fluxes were constrained by the seeing that was attained
    during our interference filter imaging observations (namely 1.9$\arcsec$, which corresponds to $\sim$ 87 pc for the
 assumed distance to NGC 2903 of 9.4 Mpc).
  Because of the difference between seeing and pixel scales, we did not include radii calculations for the 
SNR candidates that we detected. To correct flux values for interstellar extinction, 
 we used data from Cardelli, Clayton, and Mathis (1989).

 Using the SNR identification 
     technique described above, 5 SNR candidates were detected in NGC 2903 with [SII]/H$\alpha$ $\geq$ 0.4. Results of our present observations with
 candidates and corrected flux ratios
from the imaging analysis of these SNRs  are listed in Table 3. The measured H$\alpha$ flux of SNR4 
is smaller than the other four optically-identified SNRs by factors of approximately 2-3.
We caution that calculations involving the H$\alpha$ line may have
slight contamination from the [NII] lines.
We carried out these calculations in an environment with a limiting flux sensitivity
 level of 3.1$\times$10$^{-15}$ erg cm$^{-2}$ s$^{-1}$ for our imaging observations
of NGC 2903. This sensitivity limit was determined by choosing a structure
that has minimum magnitude in the galaxy field. Then the total counts from the same circular
aperture of SNRs used are converted to galaxy flux value using the spectroscopic standard star flux mentioned above.

Only one field was observed in NGC 2903.
     It covers a total field of 12\arcmin $\times$ 6\arcmin. 
In Figs. 2 - 3 we show 4${\arcmin}$ $\times$ 4${\arcmin}$ 
subfields of NGC 2903. No SNRs were found in the southern half of the galaxy. We note that the south field of the galaxy consists 
of mainly bright HII regions compared with the north field (Bresolin et al. 2005).
 This might be one of the reasons for the absence of SNRs in this region.
  The detected SNR distribution in NGC 2903  resembles the skewed distribution 
  of SNRs seen in NGC 2403 by Matonick et al. (1997). They also explain the absence of 
  SNRs in the north field of the image of NGC 2403 in a similar fashion.

Among the five candidates detected in NGC 2903, we were able to observe
SNR4 spectroscopically and derive a specific line ratio of [SII]/H$\alpha$
of 0.42. The optical spectrum of SNR4 is shown in Fig. 4. Line intensities relative to H$\beta$, the E$_{(B-V)}$, and H$\alpha$ intensity value for this spectrum given in Table 4. We used this spectrum to calculate line ratios 
 for  [SII]$\lambda$ 6761/[SII]$\lambda$6731 and [OIII]${\lambda}$5007 / H$\beta$. 
 We calculated electron density, $N_e$, using [SII]$\lambda$ 6761/[SII]$\lambda$6731 line ratio and 
 the Space Telescope Science Data Analysis System (STSDAS) task $nebular.temden$. 
 When an electron temperature value is given, this task calculates the electron density, 
 based on the five-level atom
approximation explained in the task.
 The line ratio of [SII]${\lambda}$6716/[SII]${\lambda}$6731 $>$ 1.46 gives the low density limit corresponding 
 to $N_e$ $\leq$ 10 cm$^{3}$ (Osterbrock 1989). 
Assuming an electron temperature of $T = 10^{4}$ K,
 the calculated $N_e$ value is 360, which is not so atypical for such galaxies 
(for example, SNR 19 in NGC 2403, Matonick et al. 1997).
For SNR4, we detected only a weak [OIII]${\lambda}$5007 / H$\beta$ 
 emission line ratio of $\sim$0.1 indicating  
an  oxygen-deficient  remnant, which corresponds to  a low propagation  
 velocity  below the limit of $\leq$ 100 km/s (Smith et al. 1993),  
which shuts off the nebular shocks. 
There are many examples of such weak line ratios with poor oxygen content in a number of nearby galaxies 
(Blair \& Long 2004; MF97).    
   
We searched for X-ray counterparts to the SNRs that were found by our optical observations.
 Twenty one X-ray point sources were determined in the position and extension of NGC 2903 in the Master X-Ray Catalog 
(http://heasarc.nasa.gov/W3Browse/all/xray.html). We found only one positional coincidence with the Master X-ray catalog
sources taking a 30$\arcsec$ positional error circle around the objects.
SNR3 falls in the error circle of 1RXS J093210.2+212947.
However, the tabulated position error of the catalog source was
18$\arcsec$, which was lower than our assumed error, once the error
circle diminished in size, there we found no correlation of our candidate
SNR3 with the source. This catalog contains data from the Position Sensitivity Proportional Counter (PSPC) onboard ROSAT   
(Rontgen Satellite) and Imaging Proportional Counter (IPC) onboard Einstein observatories.
 IPC provides an angular resolution of 
$\sim$30${\arcsec}$ (FWHM) at $\sim$1 keV  and the limiting sensitivity 
range from  $\sim$5$\times$10$^{-14}$ to
$\sim$3$\times$10$^{-12}$erg cm$^{-2}$s$^{-1}$ 
  in the 0.3 - 3.5 keV energy band (Gioia et al. 1990).  
And ROSAT PSPC minimum sensitivity lies around  a few times 10$^{-13}$ - 2$\times$10$^{-14}$ erg cm$^{-2}$
s$^{-1}$ in the energy band 0.1-2 keV (Morley et al. 2001). PSPC has provided $\sim$30${\arcsec}$ (FWHM)
on-axis angular resolution at 1 keV (Trumper, 1984). These limiting fluxes set 
 an upper limit on the luminosity of sources that would be detected in this catalog 
as a few times 10$^{38}$ erg s$^{-1}$ at the distance of 9.4 Mpc (for NGC 2903). Given that a maximum radiative X-ray luminosity of an SNR (with a shock 
temperature of 0.1 keV or above) will be a few times 
10$^{38-39}$ erg s$^{-1}$ (see Panutti, Schlegel, Lacey 2003; Schlegel $\&$ Panutti 2003; Holt et al. 2003) , 
it is only normal that there are no SNRs among
the Master X-ray Catalog sources in the vicinity of NGC 2903.
Tschoke et al. (2003) have also found 18 sources in the vicinity of NGC 2903.
We checked for any positional coincidence with these sources using an error 
circle of 30${\arcsec}$, but found no correlation. Given our results for 
SNR4, since we derive no strong [OIII] emission and 
our [OIII]${\lambda}$5007 / H$\beta$ ratio indicates that 
nebular shocks are around or below 
100 km/s (i.e., an old remnant), this yields an upper limit on the 
electron temperatures of 
about 10$^5$ K, which would greatly reduce the X-ray emission that will 
be detected from this remnant.
It is expected that X - ray imaging with Chandra, with improved angular resolution and sensitivity, 
could provide valuable improvement for SNR detections in  nearby galaxies especially for NGC 2903.    

We also checked for an overlap between our optically identified SNRs and the candidate radio SNR in NGC2903 reported by Tsai et al (2006).
 However, there is no overlap  (within  2$\arcsec$ positional accuracy) among these sources.

As noted by MF97, Braun \& Walterbos (1993) have estimated that about half of all SNe are of Type Ib/c or Type II ( that is, produced by the deaths of massive stars); in turn, only half of all of these SNe are located in regions with enough ambient density to produce a detectable SNR.
This means that, only about a quarter 
of all SN events may leave  easily detectable optical remnants. In our case we have detected five SNRs, indicating that the total number of SNe 
in NGC 2903  is $\sim$ 20, about half of which ($\sim$ 10) would leave  remnants. This number could also be taken as an upper limit 
for observations with much more improved visibility and seeing conditions. However, only about half of them ($\sim$ 5) would occur 
in easily detectable regions.  All these provide acceptable explanations for our observations. If the optically visible lifetime 
of a typical SNR is about 20,000 years (Braun et al. 1989), it would also give us an SN occurrence rate of about ~1 per 1000 yrs. When we compared this rate to MF97 SN rates, we find quite good overlap for the case of NGC 5585, which has the same number of SNRs. In the same work, for galaxies with lower and higher SNR numbers, this rate goes proportionally higher and lower.

In their analysis, MF97 also present the mode  values of the
measured  H$\alpha$ intensities for the
detected SNR samples from numerous galaxies to see evidence of any
selection effects and biases (see their Table 19). Using the galaxy distances they show the log of the mode of the
H${\alpha}$ luminosity,  L(H$\alpha$)$_{mode}$,
is larger for more distant galaxies. This means that, if the distance of galaxies increases, it is much more
difficult to detect the fainter SNRs.
In our case, we calculated that the
 L (H${\alpha}$)$_{mode}$ value could be taken as  $\sim$1.2$\times$10$^{38}$ erg s$^{-1}$ 
 (since we have detected a few SNRs, we were only 
 able to calculate an average value for  H$\alpha$ intensities and considered this as our mode value. 
This was also the practice by MF97 for galaxies with a low number of SNRs). With a distance of  
9.4 Mpc, NGC 2903 follows the same trend toward higher L(H$\alpha$)$_{mode}$ values to go with 
 greater distances.

 \begin{acknowledgements}
We thank the TUBITAK National Observatory (TUG) and Special Astrophysical Observatory (SAO) for their support with observing time and equipment.
 Also we would like to thank to  IGPP (Institute of Geophysical Planetary Physics) at UCR (University of California Riverside) for 
providing us some of the interference filters. We also thank an anonymous referee and M.E. Ozel for their valuable comments and discussions. 
\end{acknowledgements}

\newpage

 \begin{table}
\begin{minipage}[t]{\columnwidth}
\caption{Characteristics of the interference filters used in our observations}             
\label{catalog}      
\centering          
\begin{tabular}{c c c c l l l }     
\hline\hline       
Name & $\lambda$ & FWHM \\
~ & Wavelength ($\AA$) & $\AA$ \\    
\hline                    
[SII] & 6728 & 54 \\  
continuum  &6964 & 350 \\
H$\alpha$   & 6563 & 80 \\
continuum   &6446 & 123  \\
   
\hline                  
\end{tabular}
\end{minipage}
\end{table}  

\begin{table}
\caption{An observation log of imaging data for our target galaxy obtained with
RTT150 at TUG.}             
\label{table:1}      
\centering          
\begin{tabular}{c c c c c l l }     
\hline\hline       
Galaxy Name& Date & Filter & Exposure  \\
 &  &  &  (s) \\
\hline                    
NGC 2903 & 2006 March 4/5 & [SII] & 3$\times$600 \\
 &  2006 March 4/5 & H$\alpha$ & 3$\times$600  \\
 &   2006 March 4/5 & continuum-6446 & 2$\times$300 \\
 &  2006 March 4/5 & continuum-6964 & 2$\times$300  \\
 & 2006 March 5/6 & [SII] & 6$\times$600 \\
 &  2006 March 5/6 & H$\alpha$ & 6$\times$600  \\
 &   2006 March 5/6 & continuum-6446 & 4$\times$300 \\
 &  2006 March 5/6 & continuum-6964 & 4$\times$300  \\
   \hline                  
\end{tabular}
\end{table}
 
\begin{table}
\begin{minipage}[t]{\columnwidth}
\caption{New optical SNR candidates detected in NGC 2903.}             
\label{table:1}      
\centering          
\begin{tabular}{c c c c c l l }     
\hline\hline       
SNR Name & RA & DEC & [SII]/H$\alpha$ & I(H$\alpha$) \\
& (J2000.0)&(J2000.0) & &(erg cm$^{-2}$s$^{-1}$)\\  
\hline
SNR1 & 9:32:12.5 & +21:32:30 & 0.41 & 9.4E-15 \\
SNR2 & 9:32:13.7 & +21:30:48 & 0.74 & 1.7E-14\\
SNR3 & 9:32:10.7 & +21:29:19 & 0.57 & 1.6E-14 \\
SNR4 & 9:32:12.5 & +21:29:06 & 0.42 & 5.4E-15 \\
SNR5 & 9:32:11.1 & +21:28:23 & 0.53 & 1.1E-14 \\
\hline                  
\end{tabular}
\end{minipage}
\end{table}   
             
 \begin{table}
\begin{minipage}[t]{\columnwidth}
\caption{Relative line intensities and observational parameters for SNR4}             
\label{table:1}      
\centering          
\begin{tabular}{c c c c c l l }     
\hline\hline       
Line & SNR4\\
\hline
H$\beta$($\lambda$4861) & 100\\
OIII($\lambda$4959) & -\\
OIII($\lambda$5007) & 11\\
NII($\lambda$5200) & -\\
He($\lambda$5876) & -\\
OI($\lambda$6300) & -\\
OI($\lambda$6364) & -\\
NII($\lambda$6548) & 30\\
H$\alpha$($\lambda$6563) & 280\\
NII($\lambda$6583) & 98\\
SII($\lambda$6716) & 62\\
SII($\lambda$6731)  & 55\\
\hline 
E$_{(B-V)}$&        0.032\\
I(H$\alpha$ )&       5.4E-15 erg cm$^{-2}$s$^{-1}$\\
$[SII]$/H$\alpha$ &         0.42\\
\hline            
\end{tabular}
\end{minipage}
\end{table}      
           
     \begin{figure}
\centering
  \includegraphics[width=9cm]{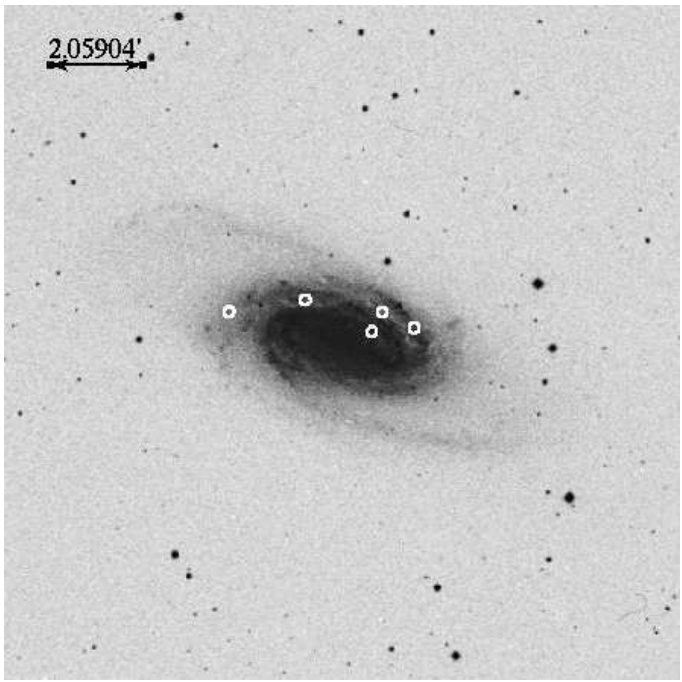}
    \caption{All the SNRs that are detected in our study are indicated on the images extracted from Digital Sky Survey (DSS).   
    Figure shows NGC 2903 with the 5 new SNR candidates found in this work labeled with circles.}
    \label{<Your label>}
\end{figure}

  \begin{figure}
   \centering
   \includegraphics[height=5cm,width=8cm]{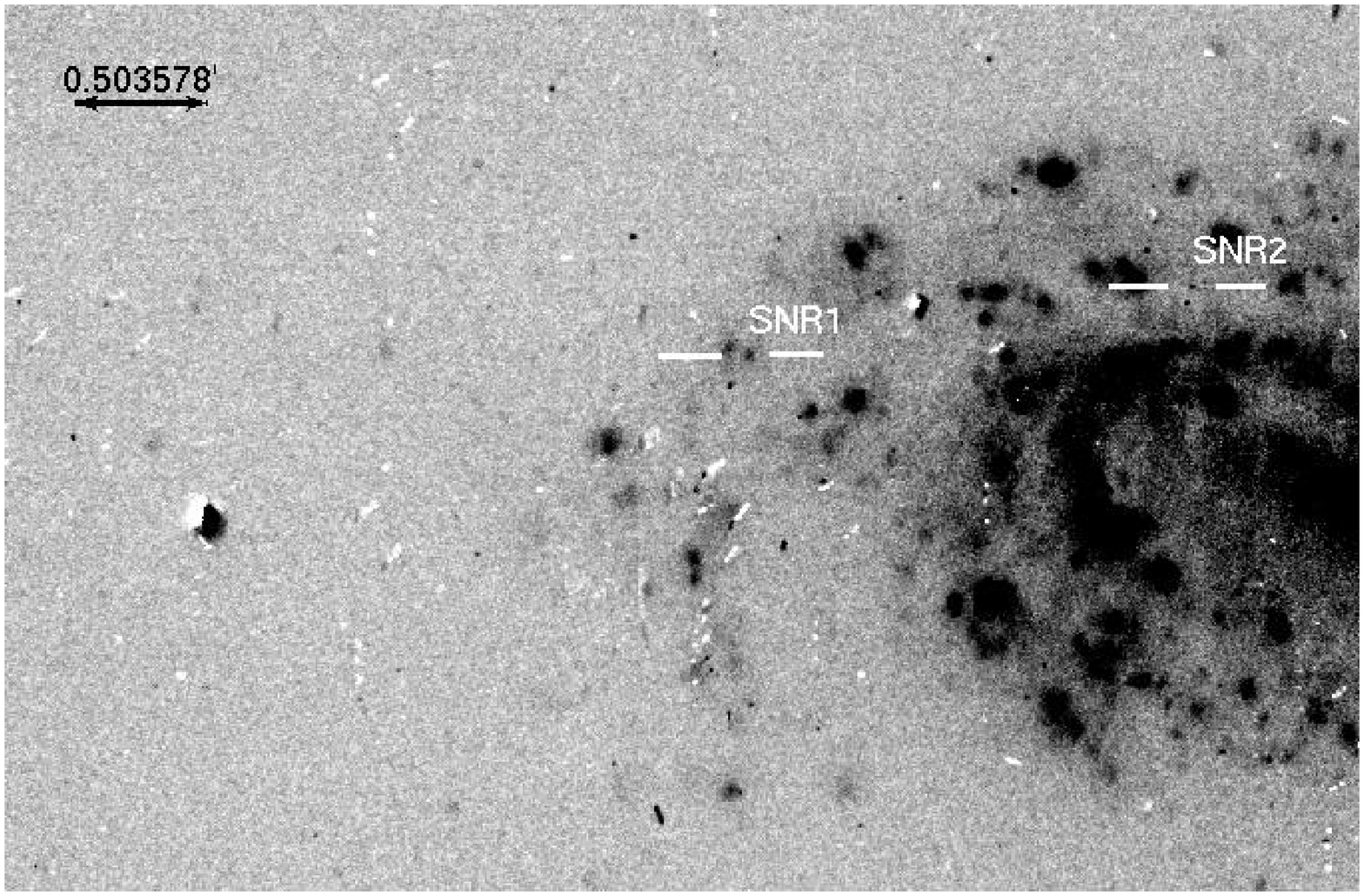}
      \caption{The figure indicates the SNR candidates (SNR1 and SNR2) detected in this work within NGC 2903 overlayed on  a $\sim$ 4 $^{\prime}$ s
subfield of continuum-subtracted [SII] image. }
         \label{FigVibStab}
   \end{figure} 

\begin{figure}
   \centering
   \includegraphics[height=5cm,width=8cm]{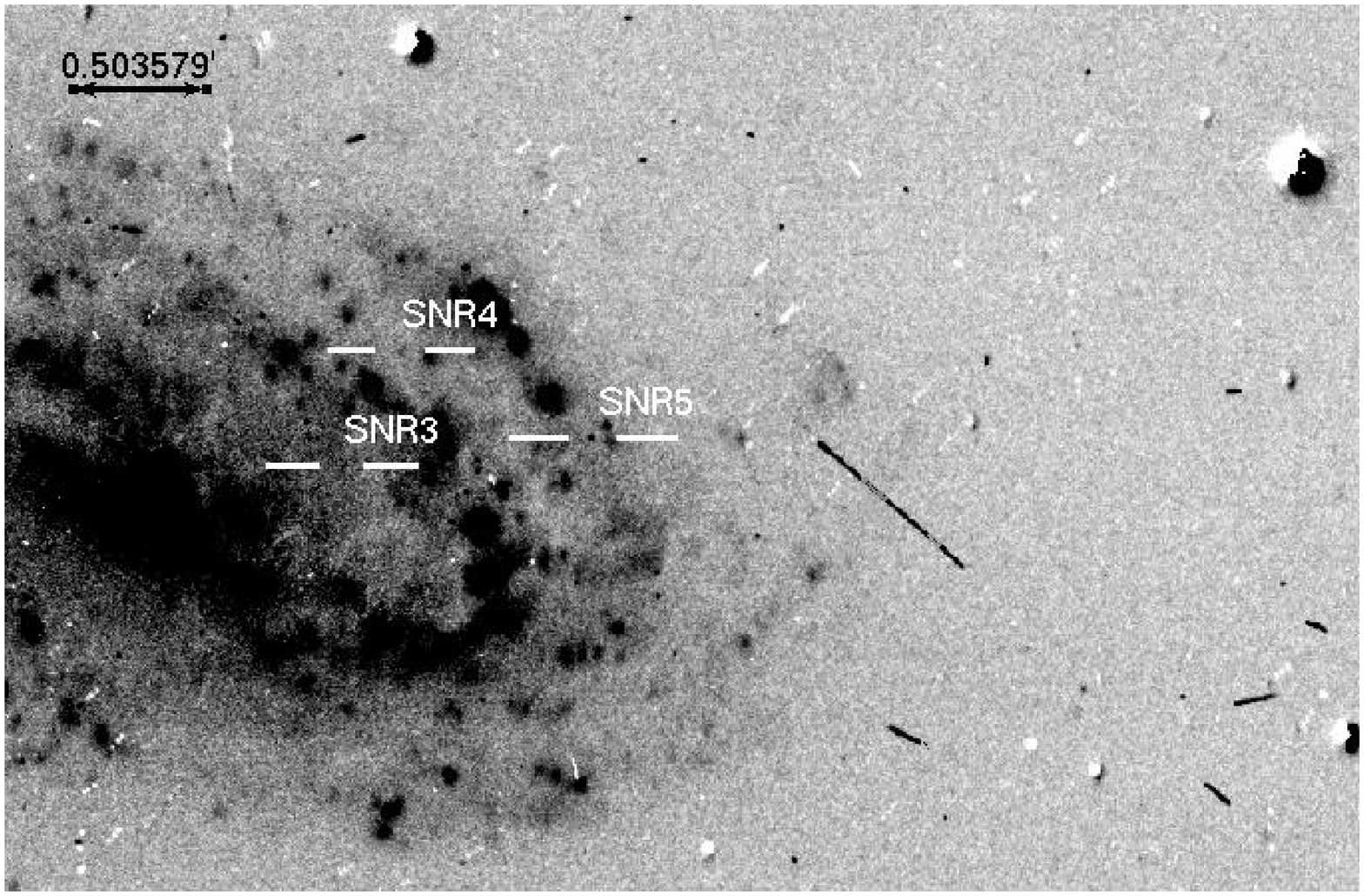}
      \caption{The figure indicates the SNR candidates (SNR3, SNR4, and SNR5) detected in this work within NGC 2903 overlayed on  a $\sim$ 4 $^{\prime}$ subfield of continuum-subtracted [SII] image. }
         \label{FigVibStab}
   \end{figure}

\begin{figure}
   \centering
\includegraphics[height=9cm,width=10cm]{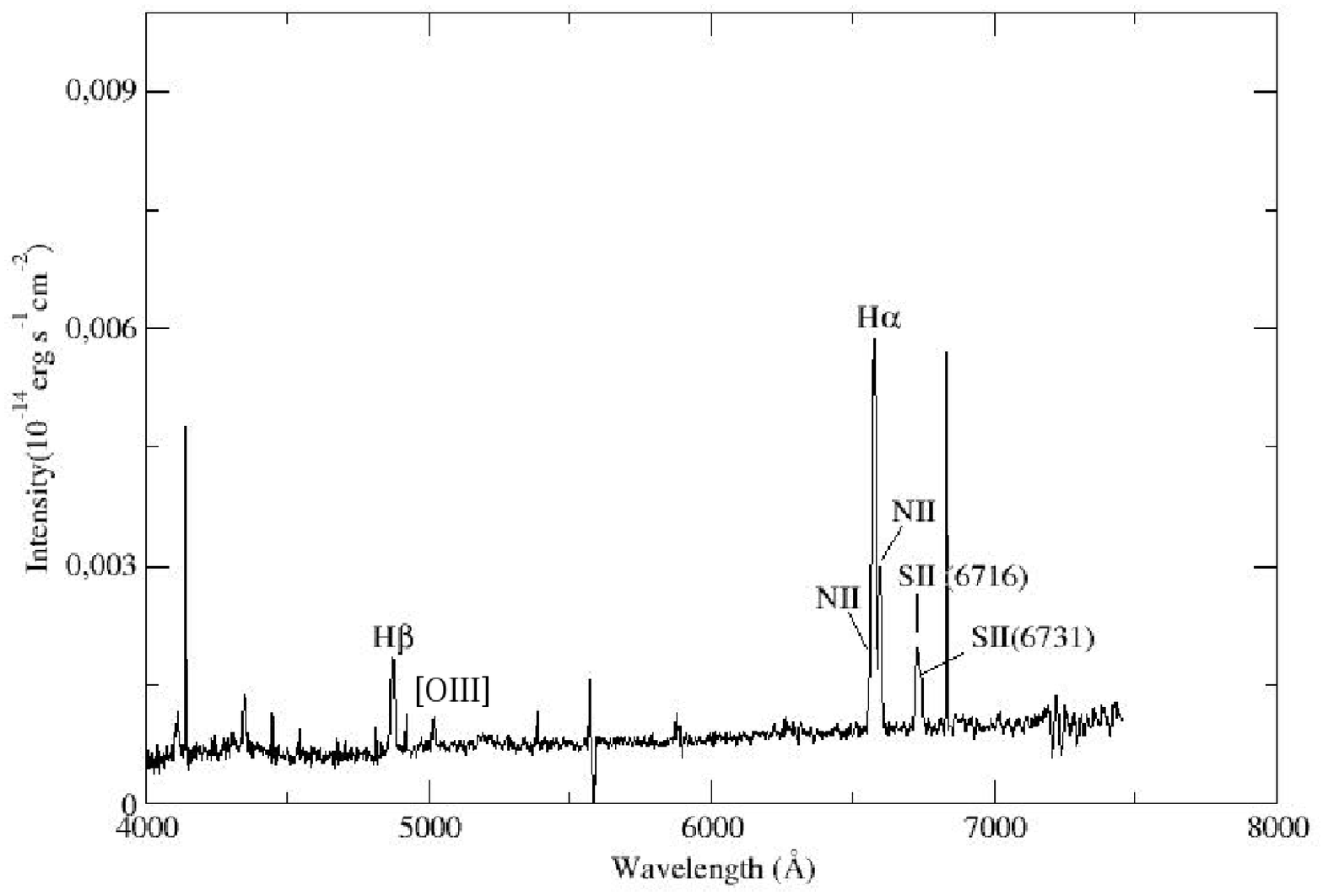}
\caption{Optical spectrum of SNR4 in NGC 2903 obtained with BTA. Identified lines are indicated in the figure.}
 \label{FigVibStab}
   \end{figure}

\end{document}